# Bank Business Models, Size, and Profitability[*]

| Fernando Bolívar | Miguel A. Duran | Ana Lozano-Vivas |
|---|---|---|
| BBVA Research (Spain) | University of Malaga (Spain) | University of Malaga (Spain) |
| fer.bolivar.garcia@gmail.com | maduran@uma.es | avivas@uma.es |

**Abstract:** To examine the relation between profitability and business models (BMs) across bank sizes, the paper proposes a research strategy based on machine learning techniques. This strategy allows for analyzing whether size and profit performance underlie BM heterogeneity, with BM identification being based on how the components of the bank portfolio contribute to profitability. The empirical exercise focuses on the European Union banking system. Our results suggest that banks with analogous levels of performance and different sizes share strategic features. Additionally, high capital ratios seem compatible with high profitability if banks, relative to their size peers, adopt a standard retail BM.

[*] The authors would like to thank Prof. Samuel Vigne, the editor-in-chief, and anonymous reviewers for their thoughtful suggestions. M. A. Duran and A. Lozano-Vivas also acknowledge financial support from the European Regional Development Fund (Project FEDERJA-169), the Andalusian Plan of Research and Development (Project P20_001010), and the Spanish Ministry for Economy and Competitiveness (Project RTI2018-097620-B-I00). F. Bolívar would also like to state that the views expressed in this paper are the authors' and do not necessarily represent those of BBVA.

# Bank Business Models, Size, and Profitability

**1. Introduction**

The effects of banks' business models (BMs) on profit performance and, therefore, on bank viability has increased banking authorities' concerns about the BM–profitability relation, especially in the European Union (EU) (European Central Bank, 2018). This paper aims to examine this relation across bank sizes to evaluate the extent to which size underlies bank strategic decisions.

In this regard, it is well known in the banking literature that size is a defining variable of banks' strategy space and, hence, conditions BMs (DeYoung and Rice, 2004). This relation between size and available strategies suggests that BMs differ across sizes, but this leaves open a question that has not been analyzed by the banking literature, specifically, whether BMs can be heterogeneous within a size or between banks of different sizes but comparable profitability. Our main contribution to the literature is our attempt to answer to this question.

Concerning the BM–profitability relation, the banking literature has adopted two main types of approach (for a review, see Badunenko et al., 2021): one based on regression analysis (e.g., Bongini et al., 2019) and another based on clustering techniques (e.g., Lagasio and Quartanta, 2022). In broad terms, the first type of approach assumes that a single variable captures the BM, thus leaving more detailed, relevant information about strategic business decisions unused. Since clustering algorithms are largely immune to standard problems in regression analysis (e.g., multicollinearity), the second type of approach can simultaneously consider the individual components of the bank portfolio to examine the BM–profitability relation. Nevertheless, earlier clustering-based works identify BMs by grouping similar banks and then computing and comparing groups' average profitability; that is, profitability is a byproduct of the BM-identifying process.

To help overcome these drawbacks of previous works, our research strategy allows us to analyze whether size influences the BM–profitability relation by simultaneously including all the components of bank portfolios and by considering profitability the basis of the BM-identifying strategy, instead of a byproduct.

The empirical exercise focuses on the EU banking system and yields two main results. First, although banks' BMs differ by size, the BMs of banks of analogous profitability share strategic similarities across sizes; that is, although the banking literature considers that size limits banks' strategy space, we find a high degree of homogeneity in how banks of different sizes but comparable profitability distinguish their BMs from those of their size peers. The second result indicates that the relation of some portfolio components with profit performance is not constant across BMs. This is the case of the capital ratio, so that the assessment of BMs by size can shed additional light on the debate about regulatory capital requirements' effects.

Indeed, the analysis provides relevant policy implications with respect to banking authorities' concerns on the BM–profitability relation. Regarding the capital ratio, our results indicate that if banks, relative to their size peers, adopt a retail BM, high profitability seems compatible with relatively high capital ratios and, hence, less risk. Thus, although the banking literature has pointed out that increasing the regulatory capital ratio reduces risk at the cost of decreasing profitability (Bank for International Settlements, 2019), our results suggest that requiring a higher capital ratio does not necessarily conflict with good profit performance, with compatibility depending on the bank BM. In a similar vein, the finding of strategic similitudes between banks of comparable profit performance and different sizes suggests that BMs can be more relevant than size in the design of profitability-enhancing policies.

## 2. Research strategy and data

Figure 1 displays the scheme of our research strategy, which proceeds in two main stages. First, since bank size can condition the available business strategies, we split our sample into large, medium-sized, and small banks. To do so, we follow the European Central Bank's criterion for the provision of consolidated banking data.[1] The second stage, which shapes the BM identification, is separately applied by size group.

**Figure 1**

As Figure 1 shows, the first step in this second stage is to build a random forest using optimal hyperparameters (Breiman, 2001). Profitability is the response variable of the random forest. In line with a portfolio-based view of BMs, the explanatory variables are banks' asset and liability portfolios' components: customer loans and deposits, interbank lending and borrowing, derivative exposures, securities, short- and long-term funding, and equity.[2] Encompassing all the portfolio components simultaneously is possible due to random forest being largely unaffected by multicollinearity, in contrast to more standard regression analysis (Zhang and Singer, 2010). Subsequently, we resort to the tree interpreter algorithm to expand the set of information that has traditionally been extracted from random forest (Kuz'min et al., 2011). In particular, as Figure 1 notes, tree interpreter allows us to compute each portfolio component's contribution to profitability at the observation level.[3]

---

[1] See https://www.ecb.europa.eu/stats/supervisory_prudential_statistics/consolidated_banking_data/html/index.en.html.

Using the portfolio components' contributions to profitability at the observation level as instruments, the third step implements the $K$-means clustering algorithm (Hartigan and Wong, 1979) to identify BMs. Accordingly, profit performance is the key variable of the BM-identifying process. This feature substantially differentiates our approach from previous clustering-based studies of the BM–profitability relation, in which profitability is viewed as a byproduct, instead of an instrument, of BM identification (Roengpitya et al., 2014; Ayadi et al., 2021; Lagasio and Quartanta, 2022).

The procedure for determining the number of BMs of size Z, with $Z = L(arge), M(edium), S(mall)$, is driven by the data. Specifically, we use the majority rule (Charrad et al., 2014) to determine the number of BMs that best fits our data in the clustering process. As Figure 1 indicates, BMλ-Z is the $\lambda$th BM of size $Z$, with $\lambda = 1, 2, \ldots, n$. For BMλ-Z, tree interpreter allows us to compute portfolio components' contributions to profitability at the observation level. Averaging these contributions by portfolio component, we can calculate the contribution of each portfolio component in BMλ-Z. Finally, adding up the contributions of each portfolio component in BMλ-Z, we obtain BMλ-Z's total contribution, $cont_{\text{BM}\lambda-Z}$, as Figure 1 shows. This overall contribution is a measure of BMλ-Z's profit performance. Therefore, we can use this type of contribution to sort the BMs within each size. In particular, λ in BMλ-Z reflects the order, that is, BM1-Z to BM*n*-Z are, for size *Z* and in terms of profit performance, the best- to worst-performing BMs, respectively.

---

[2] For the definitions of the variables capturing portfolio components, see Roengpitya et al. (2014) and Ayadi et al. (2021).
[3] To define the portfolio components' contributions in more technical terms, consider that, for an observation and a forest's single decision tree, we have the following elements: i) the tree's root node, ii) the final node where the observation is classified, iii) a path that, passing through intermediate nodes, leads from the root node to that final node, and iv) the portfolio components (i.e., the explanatory variables) chosen to split the sample at the path's nodes. Given these elements, a portfolio component's contribution to the observation's profitability measures how much the component—if selected as the splitting variable along the path—contributes to make this profitability different from the sample mean profitability. To compute the component's contribution to the observation's profitability in the entire forest, the component's contributions across the forest trees are averaged.

After the BMs are identified, the last step, labeled 4 in Figure 1, consists in defining the portfolio components characterizing them. A component characterizes BM$\lambda$-Z if two conditions are satisfied. Both conditions are based on the comparison between BM$\lambda$-Z and its complementary set, C-BM$\lambda$-Z, which is composed of banks of size *Z* not included in BM$\lambda$-Z. The first condition requires the average value of the portfolio component to be larger in BM$\lambda$-Z than in C-BM$\lambda$-Z. The second condition requires the distributions of the portfolio component in BM$\lambda$-Z and C-BM$\lambda$-Z to be different, in the sense of the Wilcoxon–Mann–Whitney test at least at the 10% significance level.[4]

Earlier works classify banks' BMs according to predefined characteristics (e.g., Caparusso et al., 2019) or consider that a single variable captures the BM (e.g., Bongini et al., 2019; Juntilla et al., 2021). By contrast, as the description above indicates, in our approach the features that characterize BMs are brought out by the BM-identifying procedure itself; that is, no ex ante assumption is made about what the identification process looks for. In this search process, since the analysis of the BM–profitability relation across sizes is aim of the paper, profit performance is, as Figure 1 shows, the element guiding what we seek.

The sample is an unbalanced panel dataset of commercial banks between 1997 and 2021 in the 15 countries belonging to the EU before its enlargement to Eastern Europe began in 2004. We collect data from the BankScope and BankFocus databases. After applying standard filters to exclude observations with meaningless values, the final dataset has 10,820 observations.

## 3. Results

Previous studies suggest that size affects the space of banks' available strategic decisions (DeYoung and Rice, 2004). Accordingly, we separately study the BM–profitability relation by size. However, to verify empirically whether the sizes' BMs actually differ, we first perform a brief between-size comparison.[5] The results of this comparison are reported in Table 1. According to the standard types of BMs described by the banking literature (retail, retail diversified, investment, and wholesale BMs),[6] large banks' BM matches the investment profile. The BMs of medium-sized and small banks share features of different standard profiles: the asset side of the retail model is noted for medium-sized banks, and the liability side for small banks, but medium-sized banks' liabilities are those of the wholesale BM, and the only relevant component of small banks' asset side is interbank lending. Thus, in line with previous research, BMs appear to differ substantially between bank sizes.

---

[4] For a detailed description of the machine learning techniques in our methodological strategy, see Bolívar et al. (2022).

[5] Defining a size's complementary set as that formed by banks of different size, the two conditions required to conclude that a portfolio component characterizes BM$\lambda$-Z are also used to establish, mutatis mutandis, whether a component characterizes a bank size's BM.

**Table 1**

According to the majority rule, used to establish the optimal number of groups to be defined by the $K$-means clustering method, we obtain three BMs per size: BM$\lambda$-Z, with $\lambda = 1, 2, 3$ and $Z = L(arge), M(edium), S(mall)$. Table 2 reports the portfolio components' contributions to profitability, multiplied by 100, by bank size and BM. The last column also shows the overall contributions of the BMs of each bank size, that is, $cont_{BM\lambda-Z}$. Within size $Z$, the BM with $\lambda = 1$, 2, or 3 has $Z$'s highest, intermediate, worst overall contribution, respectively, and, hence, it is the best, intermediate, or worst performer in terms of profitability. When we compare the BMs of banks of similar performance but different sizes, no bank size is systematically better or worse than the remaining sizes at all levels of performance. For instance, small banks with the best-performing BM, BM1-S, have the highest overall contribution (1.1160) across the sizes' BMs, but the opposite characterizes small banks with the worst-performing BM, BM3-S, since no other BM makes an overall contribution lower that theirs (-1.7668). Therefore, although size can constrain banks' strategy space, the strategic decisions of banks of the same size can give rise to substantial differences in their relative profitability.

**Table 2**

The results of applying our research strategy separately to large, medium-sized, and small banks are displayed in Table 3. Specifically, the table shows, in dark gray, the portfolio components that characterize BMs across sizes. Our analysis leads to two salient results.

Regarding the first result, the between-size comparison of the BMs above (in Table 1) indicates that large, medium-sized, and small banks' strategies differ from each other. However, as Table 3 shows, the within-size identification suggests that the BMs of banks with different sizes but comparable levels of performance share remarkable similarities. Therefore, despite the between-size differences, banks with a given degree of performance seem to differentiate from their size peers similarly across sizes.

---

[6] Customer deposits and loans and equity are the characteristic portfolio components of the retail BM, and derivative exposures, securities, and short- and long-term funding those of the investment BM. The retail diversified BM is a retail BM with short- and long-term funding, and the wholesale BM is an investment BM with interbank lending and borrowing (Roengpitya et al., 2014; Ayadi et al., 2021).

**Table 3**

In this sense, the best-performing BMs of the three sizes match (or are close to matching) a standard retail profile. This profile differs from BM1-L and BM1-M only in that short-term borrowing also characterizes these two BMs and customer deposits are not characteristic of BM1-M. Regarding small banks, BM1-S perfectly matches a standard retail profile.

We observe a similar type of outcome for large and medium-sized banks' intermediate performers. In this case, both BM2-L and BM2-M are equivalent to a standard wholesale model, except for short-term funding (derivative exposures and short-term funding) not being characteristic components of BM2-L (BM2-M). The BM of small intermediate performers is also linked to the standard wholesale profile, since interbank lending and borrowing, which are the unique components characterizing BM2-S, are not present in any standard BM except the wholesale model. Although no standard profile matches the worst-performing BM of any size, the BMs of large, medium-sized, and small banks with this level of performance are all characterized by derivative exposures.

The second salient result refers to equity. In the box of tools of regulatory authorities, a classical measure for enhancing the resilience of the banking sector is the modification of capital requirements; indeed, increasing regulatory capital is among the Basel III Accord's reforms. Despite the positive effect that this measure can have on the probability of bankruptcy, the banking literature has pointed out a side effect: holding more capital can lower profitability (Bank for International Settlements, 2019). In this regard, our analysis suggests that the BM can enrich this stability–profitability debate associated with the capital ratio. Specifically, Table 3 shows that equity is a portfolio component characterizing any size's best-performing BM—which mostly matches a standard retail profile—but it is also a characteristic component of small and medium-sized banks' worst-performing BMs; that is, the relation between capitalization and profitability seems to depend on the BM. Therefore, when the BM is included in the analysis, a relatively good profit performance target is not necessarily incompatible with being highly capitalized in relation to size peers and, hence, with greater stability.

Specifically, there does not appear to be any profitability–stability conflict if banks adopt the best BMs in Table 3, that is, if banks' BMs match or approach a standard retail profile.

## 4. Conclusion

To analyze the BM–profitability relation across sizes, this paper proposes a BM-identifying strategy that, combining machine learning techniques, uses profit performance as the main identification instrument.

The analysis offers practical insights that can be relevant to the design of banking policy in different dimensions. If the target is to improve profitability, our proposal provides a tool that can help policy makers to obtain better knowledge about the strategic decisions—at the level of either the BM or its components—that contribute to improve or worsen profit performance and which should, hence, be stimulated or penalized. Regarding the EU banking system, the paper's findings indicate that incentivizing banks to get closer to the retail BM relative to their size peers and reducing derivative exposition can be profitability-enhancing measures, regardless of bank size. Our results about equity can be of special concern for policy design. In this regard, the banking literature has pointed out that promoting banks to have high capital ratios, in line with the Basel III Agreement, is expected to reduce bank risk across sizes and BMs, but at the cost of lower profitability. Nevertheless, our outcomes suggest that high capital ratios can be compatible with high profitability, although only if we take into account BMs. In particular, improving bank stability and profitability are compatible targets if banks adopt the retail BM. This effect, once again, is observed for the three bank sizes.

Actually, the within-size comparison of BMs shows that, overall, banks of different sizes and analogous profitability share significant strategic similarities, whereas banks of the same size and unequal profit performance are strategically different. This relevant finding is not inconsistent with size limiting banks' strategy space, as the banking literature has indicated; indeed, we observe that each size's entire BM differs from those of other sizes. Nevertheless, such a finding suggests that, when banks' BMs are identified relative to their size peers, banks' strategic decisions are more important than size in the design of profitability-enhancing policies.

**Figure 1. Methodological strategy**

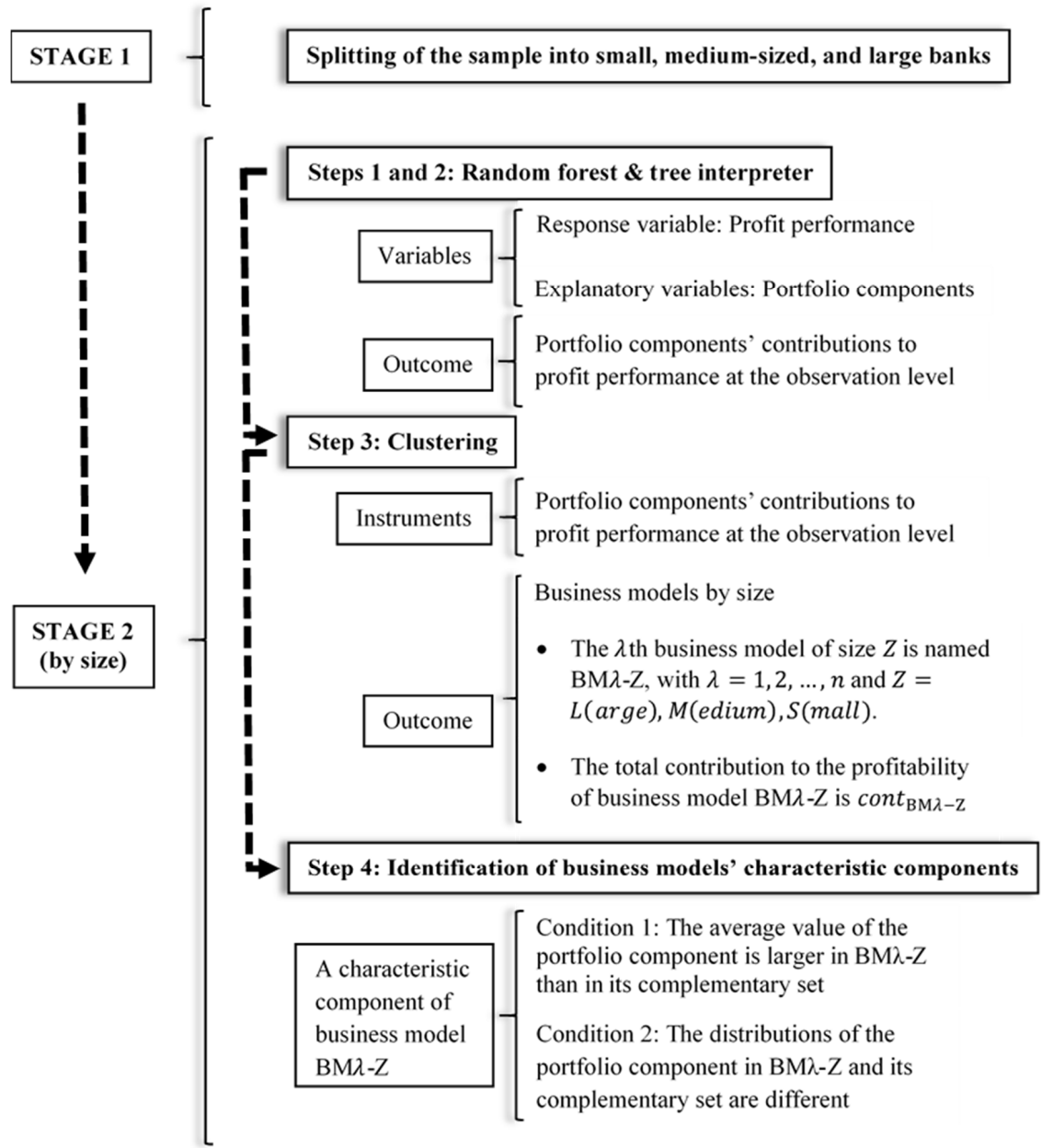

This figure synthesizes the paper's research strategy.

| | Obs. | Customer loans | Interbank lending | Derivative exposures | Securities | Customer deposits | Interbank borrowing | Short-term funding | Long-term funding | Equity |
|---|---|---|---|---|---|---|---|---|---|---|
| **Large** | 1038 | 0.4821*** | 0.1417*** | 0.1231*** | 0.2178*** | 0.4121*** | 0.1752*** | 0.0433*** | 0.1311*** | 0.0485*** |
| **C-Large** | 9782 | 0.5270 | 0.2352 | 0.0101 | 0.1548 | 0.5608 | 0.2073 | 0.0173 | 0.0649 | 0.0977 |
| **Medium** | 5921 | 0.5456*** | 0.2150*** | 0.0161*** | 0.1534*** | 0.5331*** | 0.2203*** | 0.0230*** | 0.0832*** | 0.0788*** |
| **C-Medium** | 4899 | 0.4951 | 0.2398 | 0.0269 | 0.1699 | 0.5628 | 0.1848 | 0.0159 | 0.0568 | 0.1100 |
| **Small** | 3861 | 0.4986*** | 0.2661*** | 0.0010*** | 0.1570*** | 0.6033*** | 0.1874*** | 0.0086*** | 0.0369*** | 0.1265*** |
| **C-Small** | 6959 | 0.5361 | 0.2041 | 0.0320 | 0.1630 | 0.5151 | 0.2135 | 0.0260 | 0.0904 | 0.0743 |

**Table 1. Between-size comparison**

This table shows the averages of the portfolio components by size, along with similar averages for the sizes' complementary sets (C-Large to C-Small). The Wilcoxon–Mann–Whitney test is used to compare the distributions of the components for each size and their complements. The 10%, 5%, and 1% levels of statistical significance in the results of these tests are denoted by *, **, and ***, respectively. The cell of a size's component is dark gray if the component characterizes the size's BM.

# Table 2. Contributions to profitability

## Panel A: Large banks

| BMs | Obs. | Customer loans | Interbank lending | Derivative exposures | Securities | Customer deposits | Interbank borrowing | Short-term funding | Long-term funding | Equity | Total contrib. |
|---|---|---|---|---|---|---|---|---|---|---|---|
| BM1-L | 311 | 0.0514 | 0.0333 | 0.0526 | 0.0300 | 0.0670 | 0.0447 | 0.0544 | 0.0393 | 0.0908 | 0.4635 |
| BM2-L | 671 | -0.0142 | -0.0019 | -0.0099 | -0.0038 | -0.0215 | -0.0084 | -0.0187 | -0.0103 | -0.0280 | -0.1167 |
| BM3-L | 56 | -0.1154 | -0.1617 | -0.1733 | -0.1211 | -0.1141 | -0.1477 | -0.0782 | -0.0949 | -0.1695 | -1.1760 |

## Panel B: Medium-sized banks

| BMs | Obs. | Customer loans | Interbank lending | Derivative exposures | Securities | Customer deposits | Interbank borrowing | Short-term funding | Long-term funding | Equity | Total contrib. |
|---|---|---|---|---|---|---|---|---|---|---|---|
| BM1-M | 1068 | 0.1368 | 0.1121 | 0.0576 | 0.1281 | 0.1184 | 0.1168 | 0.0506 | 0.1037 | 0.2453 | 1.0694 |
| BM2-M | 4315 | -0.0152 | -0.0043 | 0.0017 | -0.0132 | -0.0115 | -0.0101 | -0.0050 | -0.0080 | -0.0526 | -0.1183 |
| BM3-M | 538 | -0.1493 | -0.1877 | -0.1284 | -0.1485 | -0.1429 | -0.1506 | -0.0601 | -0.1420 | -0.0648 | -1.1743 |

## Panel C: Small banks

| BMs | Obs. | Customer loans | Interbank lending | Derivative exposures | Securities | Customer deposits | Interbank borrowing | Short-term funding | Long-term funding | Equity | Total contrib. |
|---|---|---|---|---|---|---|---|---|---|---|---|
| BM1-S | 1071 | 0.1414 | 0.1326 | 0.0276 | 0.1469 | 0.1523 | 0.1555 | 0.0318 | 0.1021 | 0.2257 | 1.1160 |
| BM2-S | 2449 | -0.0226 | -0.0231 | 0.0011 | -0.0297 | -0.0320 | -0.0397 | -0.0066 | -0.0217 | -0.0676 | -0.2420 |
| BM3-S | 341 | -0.2820 | -0.2503 | -0.0944 | -0.2478 | -0.2484 | -0.2029 | -0.0528 | -0.1648 | -0.2234 | -1.7668 |

For each BM$\lambda$-Z, with $\lambda = 1,2,3$ and $Z = L(arge), M(edium), S(mall)$, this table reports the portfolio components' average contributions to profitability, multiplied by 100, and, in the last column, BM$\lambda$-Z's overall contribution to profitability, that is, $cont_{BM\lambda-Z}$.

| | | | | | Table 3. BMs by size | | | | |
|---|---|---|---|---|---|---|---|---|---|---|
| **Panel A: Large banks** | | | | | | | | | | |
| BMs | Obs. | Customer loans | Interbank lending | Derivative exposures | Securities | Customer deposits | Interbank borrowing | Short-term funding | Long-term funding | Equity |
| **BM1-L** | 311 | 0.5626*** | 0.1223*** | 0.0653*** | 0.1867*** | 0.4838*** | 0.1472*** | 0.0615*** | 0.1177** | 0.0586*** |
| **C-BM1-L** | 727 | 0.4477 | 0.1501 | 0.1478 | 0.2312 | 0.3814 | 0.1871 | 0.0355 | 0.1368 | 0.0442 |
| **BM2-L** | 671 | 0.4422*** | 0.1524*** | 0.1481*** | 0.2346*** | 0.3785*** | 0.1887*** | 0.0365*** | 0.1358 | 0.0434*** |
| **C-BM2-L** | 367 | 0.5552 | 0.1223 | 0.0774 | 0.1872 | 0.4736 | 0.1503 | 0.0556 | 0.1225 | 0.0578 |
| **BM3-L** | 56 | 0.5140* | 0.1225*** | 0.1447** | 0.1899** | 0.4169 | 0.1678 | 0.0227** | 0.1493** | 0.0537 |
| **C-BM3-L** | 982 | 0.4803 | 0.1428 | 0.1219 | 0.2194 | 0.4118 | 0.1756 | 0.0445 | 0.1301 | 0.0482 |
| **Panel B: Medium-sized banks** | | | | | | | | | | |
| **BM1-M** | 1068 | 0.5764*** | 0.1987*** | 0.0137*** | 0.1377*** | 0.5235 | 0.1963*** | 0.0309*** | 0.0788*** | 0.1023*** |
| **C-BM1-M** | 4853 | 0.5388 | 0.2186 | 0.0166 | 0.1568 | 0.5353 | 0.2255 | 0.0213 | 0.0842 | 0.0737 |
| **BM2-M** | 4315 | 0.5345*** | 0.2254*** | 0.0155*** | 0.1592*** | 0.5354 | 0.2270*** | 0.0217** | 0.0853*** | 0.0709*** |
| **C-BM2-M** | 1606 | 0.5754 | 0.1870 | 0.0175 | 0.1378 | 0.5271 | 0.2023 | 0.0266 | 0.0776 | 0.1002 |
| **BM3-M** | 538 | 0.5733** | 0.1638*** | 0.0251*** | 0.1381* | 0.5342 | 0.2141 | 0.0180** | 0.0754 | 0.0958*** |
| **C-BM3-M** | 5383 | 0.5428 | 0.2201 | 0.0151 | 0.1549 | 0.5330 | 0.2209 | 0.0235 | 0.0840 | 0.0771 |
| **Panel C: Small banks** | | | | | | | | | | |
| **BM1-S** | 1071 | 0.5152*** | 0.2454*** | 0.0005*** | 0.1593 | 0.6194*** | 0.1453*** | 0.0059 | 0.0276*** | 0.1513*** |
| **C-BM1-S** | 2790 | 0.4922 | 0.2741 | 0.0012 | 0.1561 | 0.5971 | 0.2036 | 0.0096 | 0.0404 | 0.1170 |
| **BM2-S** | 2449 | 0.4911*** | 0.2791*** | 0.0011 | 0.1558 | 0.5982** | 0.2080*** | 0.0096 | 0.0389 | 0.1141*** |
| **C-BM2-S** | 1412 | 0.5115 | 0.2436 | 0.0008 | 0.1591 | 0.6121 | 0.1518 | 0.0067 | 0.0334 | 0.1481 |
| **BM3-S** | 341 | 0.4999 | 0.2377** | 0.0016*** | 0.1585 | 0.5892* | 0.1722 | 0.0091 | 0.0515*** | 0.1381*** |
| **C-BM3-S** | 3520 | 0.4985 | 0.2689 | 0.0009 | 0.1568 | 0.6047 | 0.1889 | 0.0085 | 0.0355 | 0.1254 |

This table reports the averages of the portfolio components in BMλ-Z and their complementary sets, C-BMλ-Z, with $\lambda = 1, 2, 3$, and $Z = L(arge), M(edium), S(mall)$. For a given $\lambda$ and $Z$, the Wilcoxon–Mann–Whitney test is used to compare the distributions of the components in BMλ-Z and C-BMλ-Z. The 10%, 5%, and 1% levels of statistical significance in the results of these tests are denoted by *, **, and ***, respectively. The cell of a component of BMλ-Z is dark gray if the component characterizes BMλ-Z.